# Coherent control of injection currents in high-quality films of Bi$_2$Se$_3$

D. A. Bas, K. Vargas-Velez, S. Babakiray, T. A. Johnson, P. Borisov,

T. D. Stanescu, D. Lederman, and A. D. Bristow*

Department of Physics and Astronomy, West Virginia University, Morgantown, West Virginia 26506-6315, USA

*E-mail: alan.bristow@mail.wvu.edu

**Abstract**

Films of the topological insulator Bi$_2$Se$_3$ are grown by molecular beam epitaxy with *in-situ* reflection high-energy electron diffraction. The films are shown to be high-quality by X-ray reflectivity and diffraction and atomic-force microscopy. Quantum interference control of photocurrents is observed by excitation with harmonically related pulses and detected by terahertz radiation. The injection current obeys the expected excitation irradiance dependence, showing linear dependence on the fundamental pulse irradiance and square-root irradiance dependence of the frequency-doubled optical pulses. The injection current also follows a sinusoidal relative-phase dependence between the two excitation pulses. These results confirm the third-order nonlinear optical origins of the coherently controlled injection current. Experiments are compared to a tight-binding band structure to illustrate the possible optical transitions that occur in creating the injection current.

$Bi_2Se_3$ is a well-known thermoelectric material[1] and currently of great interest in condensed matter physics because the surface states exhibit massless Dirac dispersions in the gaps between the bulk valence and conduction bands.[2] The surface states have an electron spin structure defined by strong spin-orbit coupling which locks the spin vector perpendicular to the momentum vector.[3,4] Fundamental properties of $Bi_2Se_3$ have been widely investigated because of the potential opportunities for developing spintronic and optically controlled devices.

A great deal of attention has been paid to the light-matter interactions because the surface states of $Bi_2Se_3$ are obscured in electrical measurements, due to inherent n-doping in the as-grown material. This means that the Fermi level is in the bulk conduction band above the 0.3-eV band gap. Despite this fact, optical measurements have been widely explored in the near infrared, revealing anomalous absorption coefficients with film thickness,[5] strong coupling between electrons and phonons[6,7] and a second set of massless surface states at about 1.7 eV above the commonly studied Dirac point below the Fermi level.[8] These developments, among others,[9] indicate that near-infrared optical interactions provide mechanisms that can be exploited for photonic devices and also access surface-surface state excitations. In which case, coupling of optical radiation to inject and control photocurrents is then relevant for the performance of this material system and coupling to the all-important surface states.

Photocurrents in $Bi_2Se_3$ have been predicted[10] and measured[11] using circular polarization, through the circular photogalvanic effect. Experimental studies have been performed in exfoliated $Bi_2Se_3$ patterned with contacts that are sensitive to both photocurrents and photo-thermal currents. Nonetheless, this study has demonstrated that photocurrents can indeed be controlled. A more discrete scheme has been proposed for coherent control of the surface states using two-color quantum interference.[12] Quantum interference uses harmonically related pulses to link the same initial and final states via single and two-photon absorption, such that it is a quantum analogue of the classical Young's double slit experiment. The two-color proposal is based on undoped $Bi_2Se_3$ where an injection current is created by quantum interference control of carriers being promoted from below the Dirac point to above it. In order to achieve this experimentally, mid-infrared light pulses would be required to impinge as-grown $Bi_2Se_3$ devices with adjustable Fermi levels, through precise gate-voltage control. Alternatively, one could exploit surface-state to surface-state transitions from the well-known Dirac cone to the newly discovered Dirac cone at higher energies. To date, these transitions have generally not been factored into many of the optical measurements performed with commercially available near-infrared lasers, despite the propensity with which they have been used. One exception is the investigation of carrier dynamics by differential reflection.[13]

Here we demonstrate the use of the two-color excitation scheme[14] to inject ballistic photocurrents in high-quality $Bi_2Se_3$ films. We also show that the currents can be controlled by the relative phase between the fundamental and frequency doubled excitation pulses. Moreover, we confirm that the injection current follows the expected dependences on irradiance of the two excitation pulses.

$Bi_2Se_3$ films of various thickness were grown by molecular beam epitaxy[15] on double-side polished 0.5-mm $Al_2O_3(0001)$ substrates by a two-step process.[16] In the first step, three quintuple layers (QLs) were grown at 140 $^\circ$C to help deposition of the film with the correct stoichiometry. The remainder of the film is grown in a second step at a temperature of 275 $^\circ$C to produce a smoother film. Figure 1(a) shows the oscillations of layer growth during the high-temperature growth stage, monitored by the *in-situ* reflection high-energy electron diffraction (RHEED). The inset of the figure shows the striping associated with a high-quality film. RHEED oscillations confirm growth of the 12 additional QLs of $Bi_2Se_3$ to form a 15-QL thick sample.

It is suspected that selenium vacancies[17,18] quickly alter the surface potential and subsequently lead to oxidation when exposed to air. Consequently, a 10-nm thick $MgF_2$ capping layer is sputtered onto the samples without removing them from the growth chamber. The result is an encapsulated and highly stable film of $Bi_2Se_3$. A schematic diagram of the sample structure is shown in the inset of Fig. 2.

$Bi_2Se_3$ film thicknesses were then confirmed with X-ray reflectivity (XRR); see Fig 1(b). Fitting the XRR pattern indicates roughness of 0.2 nm. Fitting of the XRR data is performed using the open-source program GenX. The inset shows an atomic force micrograph of the $Bi_2Se_3$ film with $MgF_2$ layer. The triangles reveal the expected degree of dislocations in the $Bi_2Se_3$ film. Figure 1(c) shows X-ray diffraction (XRD), which is used to determine the degree of the out-of-plane disorder. The 003 XRD Bragg peak observed at low angle corresponds to that observed in the XRR data. Overall, characterization indicates high-quality films with which to perform optical measurements.

Figure 2 shows the experimental setup for the current injection measurements. A laser amplifier system with optical parametric amplifier (OPA) provides ~80-fs pulses at a repletion rate of 1 kHz. Signal pulses from the OPA are centered at 1540 nm (0.8 eV). A β-barium borate crystal frequency doubles the fundamental pulses, generating second-harmonic pulses at 770 nm (1.6 eV). These fundamental ($\omega$) and frequency-doubled ($2\omega$) pulses feed a two-color Mach-Zehnder interferometer that allows for independent control of the phase and polarization for each color. The two colors are set to collinear polarization and their time delay is adjusted to be close to the center of their envelopes, as measured with an interferometric cross-correlation. Measurements presented here are performed on the 15-QL thick $Bi_2Se_3$ sample at normal incidence. Measurements have also been performed on samples ranging from 6 QL to 40 QL showing qualitatively consistent behavior.

The acceleration of charge in the production of the photocurrent on a picosecond timescale re-radiates light at terahertz (THz) frequencies.[19,20] The sample is oriented such that the optical pulses transmit through the sapphire substrate before they arrive at the $Bi_2Se_3$ film, so as to not distort the emitted THz radiation.[21] The THz radiation is collected using off-axis parabolic mirrors and focused onto a 1-mm thick electro-optic crystal (ZnTe), which is gated by an 80-fs pulse from the laser amplifier. The quasi-static electric field of the THz induces birefringence in the ZnTe that is seen by the gate pulse and measured by projecting the orthogonal polarization components onto balanced photodetectors.[22] The signal is extracted using lock-in detection referenced to a mechanical chopper placed close to the laser source, which modulates the light at 250 Hz. The THz signal is mapped out in the time domain by varying the relative time delay between the emitted THz and the gate pulse.

Figure 3(a) shows THz transients recorded for two excitation conditions of the relative phase ($\Delta\phi = 2\phi_\omega - \phi_{2\omega}$) between the $\omega$ and $2\omega$ optical pulses. It can be seen that for an excitation condition of $\Delta\phi =$ $3\pi/2$ the transient has the opposite sign to that for $\Delta\phi = \pi/2$. Figure 3(b) shows a sinusoidal dependence of the THz amplitude versus $\Delta\phi$, measured at the peak of the THz transient indicated by the dashed vertical line in Fig. 3(a). In both Fig. 3(b) a slight offset from zero is observed indicating that there exists a weak background that is phase independent and results from excitation due to only one pulse.

Figure 3(c) shows the dependence of the emitted THz amplitude on irradiance (or average power) of one excitation pulse while holding the other constant and vice versa. Fixing the average power of the $\omega$ pulse and varying the average power of the $2\omega$ pulse results in a sub-linear slope on the log-log plot, with a $\sqrt{I_{2\omega}}$ dependence. In this case, the THz amplitude scales because of single-photon absorption of the $2\omega$ pulse. In contrast, fixing the average power of the $2\omega$ pulse and varying the average power of the $\omega$ pulse results in a linear slope, with an $I_\omega$ dependence. In this case, the THz amplitude scales due to two-photon absorption of the $\omega$ pulse. Overall, the emitted field amplitude of the THz signal follows a $\sqrt{I^3}$ (or $I^{1.5}$) dependence, confirming that it is governed by a third-order nonlinear optical process.

The relative-phase dependence and average-power dependence of the emitted THz signals are the expected signatures of a signal derived from an injection current.[23] Injection currents arise from an imbalance of the total number of carriers excited to positive and negative states in *k*-space. The net momentum is equivalent to excitation of a ballistic photocurrent. In this all-optical method of injecting charge currents the emitted THz follows the derivative of the injection current $J_i$, such that $dJ_i/dt = 2\eta_{ijkl}(0;\omega,\omega,-2\omega)|E^\omega_{jk}|^2 E^{2\omega}_l \sin\Delta\phi$, where $\eta_{ijkl}$ is a fourth-rank tensor related to the imaginary part of $\chi^{(3)}$, $E^\omega_{jk}$ is the $\omega$ electric field, $E^{2\omega}_l$ is the $2\omega$ electric field and $i, j, k$ and $l$ are numerical indices.[24] Using the above equation the relative-phase dependence and average power-dependences have been fit, shown as solid lines in Fig. 3(b) and (c) respectively. Overall the theoretical comparison qualitatively fits the experimental data well. At present, quantitative analysis cannot be performed because little is known about the nonlinear absorption coefficients for this material due to saturable absorption.[9] Nonetheless, theoretical verification of the irradiance and relative phase dependence exclude other THz generation processes such as optical rectification[22] and shift currents.[20]

Third-order nonlinear processes are expected from all materials and injection currents have been observed in a range of solids, including unbiased and unstrained silicon,[14] direct-gap GaAs,[19] nanowires[25] and graphene.[26] The process in $Bi_2Se_3$ may be similar to that in graphene if Dirac surface states are involved. In which case, quantum interference between single and two-photon absorption at $2\omega$ occurs in the presence of simultaneous single-photon absorption of the $\omega$ pulse. In graphene, the absorption at $\omega$ does not affect the injection current, as would be expected in this case as well.

The inset of Fig. 2 shows a band structure for a 15-QL $Bi_2Se_3$ thin film calculated using a phenomenological tight-binding model, based on eight bands to capture both sets of surface states and Dirac cones. The parameters are optimized for a bulk system using results from local-density approximation calculations[27,28] and two-photon photoemission measurements.[8] Confinement-induced bands are determined by solving the model in the slab geometry. The $2\omega$ frequency is also illustrated and is commensurate with excitations from the surface states near the $\Gamma$-point of the band structure below the Fermi surface to states in the newly-discovered set of surface states at close to 1.7 eV.

At the wavelength of excitation, the transitions depend on the character of the bulk and surface states. There are a varierty of transitions that include surface, bulk and both types of states. Indeed, the surface-surface transitions and bulk-bulk transitions should have very different characteristics. Moreover, bulk-surface or surface-bulk transitions lead to a charge displacement in the thin films. Finally, in thin films confinement can lead to new bands in the band structure drastically changing the ratio of surface and bulk states suitable for optical transitions.[29]

In summary, $Bi_2Se_3$ films have been grown to sufficient quality that injection currents have been observed and controlled using all-optical methods. All-optical methods of charge current excitation and detection are non-invasive and avoid problems with contacts, such as local modification of band structures and slow read-outs that can confuse current signals. The ability to inject and control these photocurrents offers a powerful method for characterization of topological insulator materials based on as-grown $Bi_2Se_3$.

**Acknowledgements:** The authors wish to thank John Sipe and Rodrigo Muniz for useful discussions. This work was supported by the West Virginia Higher Education Policy Commission (HEPC.dsr.12.29) and WVU Shared Research Facilities. KVV was supported by the National Science Foundation through the Research Experience for Undergraduates program (DMR-1262075).

**Figures**

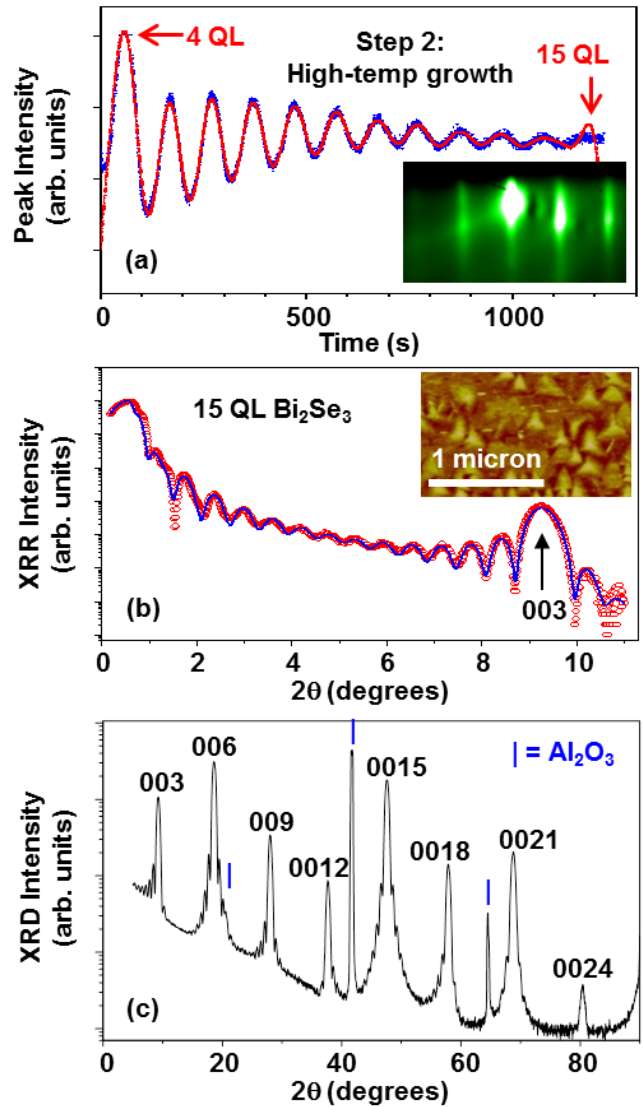

Figure 1 (a) Reflection high-energy electron diffraction during the second growth stage of the Bi$_2$Se$_3$ film. The oscillations indicate growth of 12 quintuple layers (QL) in the formation of a 15-QL sample. [1 QL is approximately 1 nm]. The inset shows clean stripes indicating good sample quality. (b) X-ray reflection of the 15-QL sample. The inset shows an atomic force micrograph of the sample. (c) Out-of-plane X-ray diffraction showing features from the substrate and Bi$_2$Se$_3$ film.

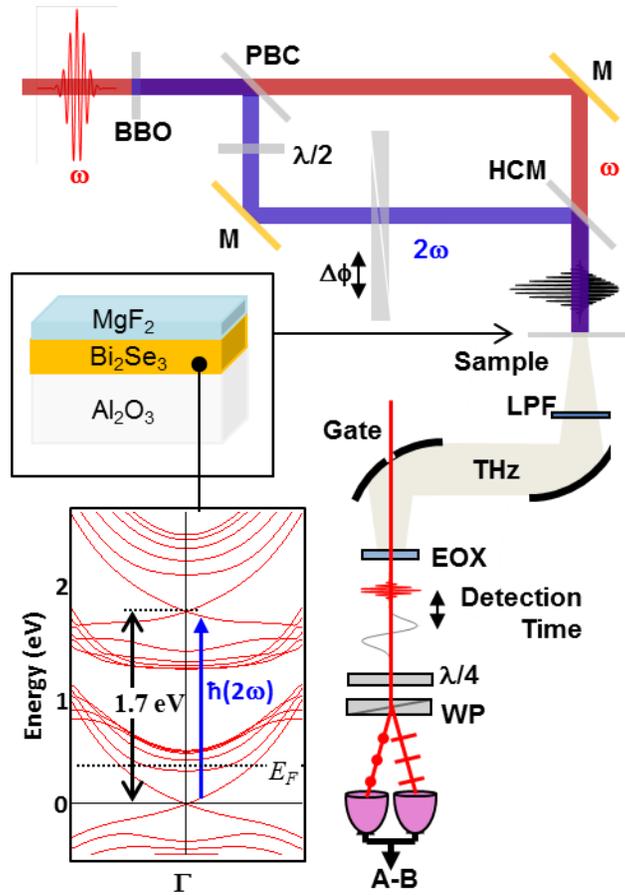

Figure 2 Experimental setup of the all-optical injection current excitation and detection scheme. BBO = beta-barium borate crystal, PBC = polarization beam cube, λ/2 = half-wave plate, M = mirror, Δϕ = relative phase between fundamental (ω) and second-harmonic (2ω) pulses, HCM = hot-cold mirror, φ = azimuthal angle, LPF = low-pass filter, EOX = electro-optic sampling crystal, λ/4 = quarter-wave plate, WP = Wollaston prism, A-B = balanced detection signal feed to lock-in amplifier. The top inset shows the structure of the sample. The lower inset shows the band structure for a slab of $Bi_2Se_3$.

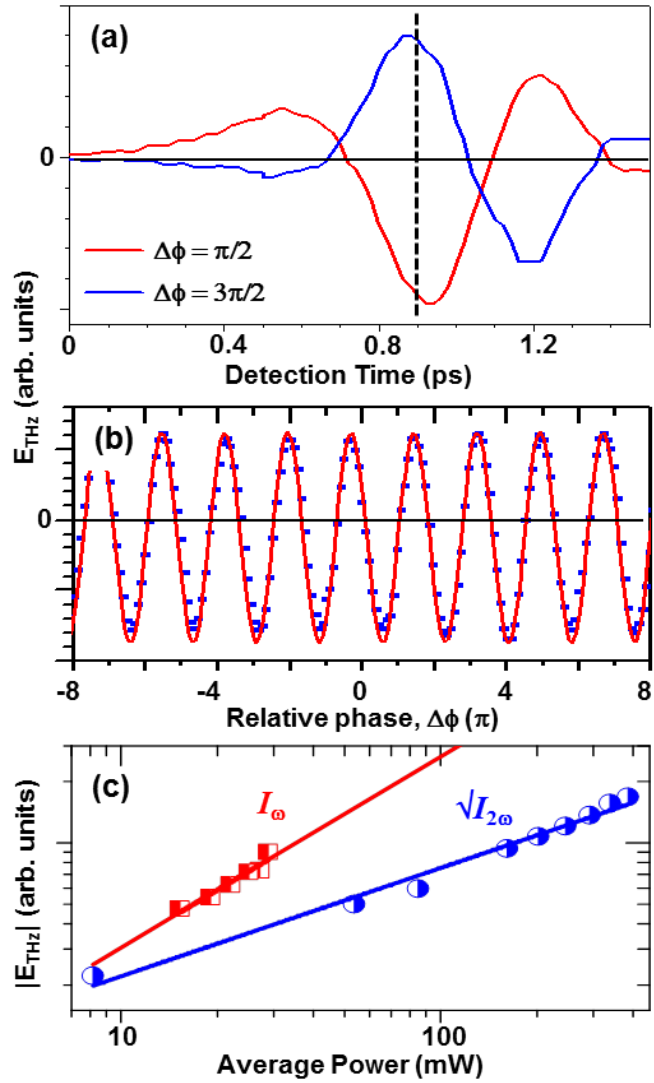

Figure 3 (a) Terahertz (THz) transients recorded with relative phase conditions between the ω and 2ω excitation pulses of π/2 and 3π/2. (b) Phase dependence recorded at the peak of the THz transient indicated in (a). (c) Average power (or irradiance) dependence of the ω and 2ω excitation pulses individually, while fixing the average power (or irradiance) of the other.